\documentclass[journal=jacsat,manuscript=article]{achemso}

\usepackage[version=3]{mhchem} % Formula subscripts using \ce{}
\usepackage{xspace}
\usepackage{url}
\usepackage{hyperref}
\usepackage{glossaries}

\newacronym{spr}{SPR}{surface plasmon resonance}
\newacronym{sers}{SERS}{surface-enhanced Raman scattering}
\newacronym{lspr}{LSPR}{localized surface plasmon resonance}
\newacronym{nir}{NIR}{near-infrared}
\newacronym{sms}{SMS}{spatial modulation spectroscopy}
\newacronym{tem}{TEM}{transmission electron microscopy}
\newacronym{sem}{SEM}{scanning electron microscopy}
\newacronym{gnr}{GNR}{gold nanorod}
\newacronym{liob}{LIOB}{laser induced optical breakdown}
\newacronym{lib}{LIB}{laser induced breakdown}
\newacronym{litb}{LITB}{laser induced thermal breakdown}
\newacronym{mpi}{MPI}{multiphoton ionization}
\newacronym{ai}{AI}{avalanche ionization}
\newacronym{pte}{PTE}{photo-thermal emission}
\newacronym{ti}{TI}{thermal ionization}
\newacronym{fem}{FEM}{finite element method}
\newacronym{dda}{DDA}{discrete-dipole approximation}
\newacronym{bem}{BEM}{boundary element method}
\newacronym{fdtd}{FDTD}{finite difference time domain}
\newacronym{fe}{FE}{finite element}
\newacronym{fwhm}{fwhm}{full-width at half-maximum}
\newacronym{pec}{PEC}{perfect electric conductor}
\newacronym{pmc}{PMC}{perfect magnetic conductor}
\newacronym{abc}{ABC}{absorbing boundary condition}
\newacronym{pml}{PML}{perfectly matched layer}
\newacronym{ctab}{CTAB}{cetyltrimethyllammonium bromide}
\newacronym{ns}{ns}{nanosecond}
\newacronym{ps}{ps}{picosecond}
\newacronym{fs}{fs}{femtosecond}
\newacronym{pde}{PDE}{partial differential equation}
\newacronym{cem}{CEM}{computational electromagnetic methods}
\newacronym{ttm}{TTM}{two-temperature model}
\newacronym{cb}{CB}{conduction band}
\newacronym{vb}{VB}{valence band}
\newacronym{iba}{IBA}{inverse Bremsstrahlung absorption}
\newacronym{ic}{IC}{ionization cascade}
\newacronym{em}{EM}{electromagnetic}
\newacronym{ht}{HT}{heat transfer}

\author{Yevgeniy R. Davletshin}
\email{yevgeniy.davletshin@ryerson.ca}
\author{J. Carl Kumaradas}
\affiliation{Department of Physics, Ryerson University, Toronto, ON, M5B 2K3, Canada}
\title[Theoretical analysis of optoporation efficiency and bubble formation during femtosecond pulse interaction with gold nanoshell]
  {Theoretical analysis of optoporation efficiency and bubble formation during femtosecond pulse interaction with gold nanoshell}

\begin{document}

% \begin{tocentry}

% Some journals require a graphical entry for the Table of Contents.
% This should be laid out ``print ready'' so that the sizing of the

% Inside the \texttt{tocentry} environment, the font used is Helvetica
% 8\,pt, as required by \emph{Journal of the American Chemical

% The surrounding frame is 9\,cm by 3.5\,cm, which is the maximum
% permitted for  \emph{Journal of the American Chemical Society}
% graphical table of content entries. The box will not resize if the
% content is too big: instead it will overflow the edge of the box.

% This box and the associated title will always be printed on a
% separate page at the end of the document.

% \end{tocentry}

\begin{abstract}
Our group has recently developed a finite element model of a nanoparticle-mediated optical breakdown phenomena. Previously, this model was used to analyze the role of the nanoparticle morphology and the wavelength dependence of a nanoparticle-mediated optical breakdown threshold during near-infrared ps and fs pulse exposures. In this study, we provide a theoretical insight into the optoporation efficiency of live cells and bubble formation threshold during nanoparticle-mediated optical breakdown. It was done by the calculation of maximum temperature and free electron density in the vicinity of a single gold nanoshell in water during 70 femtosecond single pulse exposure and comparison against published experimental data. 
\end{abstract}

\section{Introduction}

An optical breakdown threshold is the laser irradiance needed to cause the \gls{liob} and is often used to characterize optical breakdown. Experimentally, optical breakdown in water (which can mimic the biological environment) can be associated with two phenomena: transition of the water from a liquid to a gas phase (caused by energy deposition) and luminescence by the plasma. The transition to the gas phase (cavitation and bubble formation) during optical breakdown is easily detectable. The temperature of water can be related to the threshold of bubble formation and does not depend on the laser pulse duration. On the other hand, the temperature of the plasma can be related to the threshold of luminescence, but it depends on the laser pulse duration. The brightness of the luminescence strongly depend on the plasma temperature. 

Theoretically, optical breakdown threshold is usually defined as heating of medium by the plasma to the boiling temperature (440.7~K~\cite{Linz2016femto}) or reaching a critical electron density in the medium, which is on the order of $\rho_{\rm crit}\approx10^{18}-10^{21}$~cm$^{-3}$ \cite{Vogel1996,Kennedy95_1,Sacchi91,Bloembergen74,Feng97,Noack1999,Vogel2005}. Although in most of the published theoretical studies \cite{Boulais2012,Hatef2016, Lachaine2016}, researchers defining optical breakdown threshold as a plasma density reaching a certain electron density that will lead to bubble formation or luminescence, the three orders of magnitude difference make this optical breakdown threshold criteria less reliable. On the other hand, the boiling temperature  is defined by lesser uncertainty in comparison to plasma density \gls{liob} threshold and therefore is more reliable as a computational criterion for optical breakdown threshold \cite{VogelNavy,Linz2016femto}.

Our group has recently published two studies \cite{Davletshin2016,Davletshin2016b}, where we have updated the physics of nanoparticle-mediated \gls{liob} phenomena in water and studied the role of the nanoparticle morphology and the wavelength dependence of the optical breakdown threshold during \gls{ps} and \gls{fs} laser pulse interaction with gold nanoparticles. We have shown that optical breakdown threshold, during \gls{ps} pulse exposure of plasmon coupled and uncoupled gold nanoparticles of different morphology, had a stronger dependence of the optical near-field enhancement than on the mass or absorption cross-section of the nanostructure \cite{Davletshin2016}. In another paper \cite{Davletshin2016b}, our group updated the theoretical model of \gls{liob} in accordance with the latest findings on the wavelength dependence of the band structure of water \cite{Linz2015,Linz2016femto} and studied the wavelength dependence of nanoparticle-mediated optical breakdown threshold during \gls{nir} \gls{fs} and \gls{ps} laser pulse durations.

In order to validate our wavelength dependent nanoparticle-mediated \gls{liob} model, in \cite{Davletshin2016b} we have used a recently published data by \citet{Lachaine2016} to compare the experimentally obtained bubble formation threshold \cite{Lachaine2016} and theoretically predicted one. This comparison had a good agreement against experimental threshold for bubble formation in the vicinity of a 112~nm diameter silica core and a 15~nm thick gold nanoshell (NS800). For our model we were able to deduce a maximum free electron plasma density, $\rho_{\rm e,max}=1.75\times 10^{20}~{\rm cm^{-3}}$, in the vicinity of NS800 nanoparticle that corresponded to experimental bubble formation threshold~\cite{Lachaine2016}. The $\rho_{\rm e,max}$ value was in agreement with the free electron density of a bubble formation threshold, $\rho_{\rm bf}=1.8\times 10^{20}~{\rm cm^{-3}}$, for pure water that \citet{Linz2015,Linz2016femto} had recently reported. Additionally, in the experimental data of \citet{Lachaine2016}, the bubble growth dynamics with fluence of the laser had two distinct regimes. The fluence of transition between two regimes of a bubble growth coincided with the fluence threshold of reaching a theoretical critical plasma density, $\rho_{\rm crit}$~\cite{Davletshin2016b}. Since there was no direct comparison of the temperature and free electron density as a criteria for nanoparticle-mediated \gls{liob} and due to differences in uncertainties associated with each optical breakdown criterion, it would be beneficial for the research community to compare the dynamics of temperature and the free electron density rise in the vicinity of the nanoparticle during laser pulse exposure.

Therefore, in this letter we will compare the maximum free electron density of the plasma to the maximum temperature of the water achieved during 70~fs pulse exposure at 800~nm wavelength at different fluences in the vicinity of two types gold nanoshells. Two gold nanoshell morphologies were used, NS660 and NS800 with the same dimensions as reported by \citet{Lachaine2016} to directly compare our theoretical findings against experimental data of the optoporation efficiency of live cells \cite{Lachaine2016}. This comparison will help to evaluate the appropriateness of free electron density and temperature as a criterion for the calculation of optical breakdown threshold. This comparison will help to understand how the increase of the incident laser pulse fluence will limit optoporation efficiency and why calculations of the critical electron plasma density and temperature of the water are equally important. Overall aim of this letter is to help researchers in the biomedical fields that utilize nanoparticle-mediated optical breakdown to better understand the results of their experimental findings. 
%%%%%%%%%%%%%%%%%%%%%%%%%%%%%%%%%%%%%%%%%%%%%%%%%%%%%%%%%%%%%%%%%%%%%
%% When referencing objects, LaTeX offers a label--ref mechanism.
%% The beilstein class extends this approach with the \cref command,
%% that adds the corresponding type of the object as well.
%%
%% Tables, figures and schemes must have a single column or double
%% column width. To make life easier, some commands are defined.
%%
%% Captions (legends) will always be added at the correct place no
%% matter where you put in the source code.
%% Please note: labels always have to come /after/ the \caption.
%%%%%%%%%%%%%%%%%%%%%%%%%%%%%%%%%%%%%%%%%%%%%%%%%%%%%%%%%%%%%%%%%%%%%
\section{Methods}
The \gls{fe} model of nanoparticle-mediated optical breakdown phenomena was build using commercial \gls{fe} software - COMSOL Multiphysics version 5.2, and is similar to one that is reported in our previous publications \cite{Davletshin2016,Davletshin2016b}. The interaction of the single 70~fs pulse at 800~nm wavelength was modelled with two types of gold nanoshells in water, one with a 78~nm diameter silica core and a 28~nm thick gold shell (NS660) and another with a 112~nm diameter silica core and a 15~nm thick gold shell (NS800), which had same dimension that were reported in experiments of~\citet{Lachaine2016}.

\section{Results and Discussion}

\begin{figure}
\centering
\includegraphics[width=\textwidth]{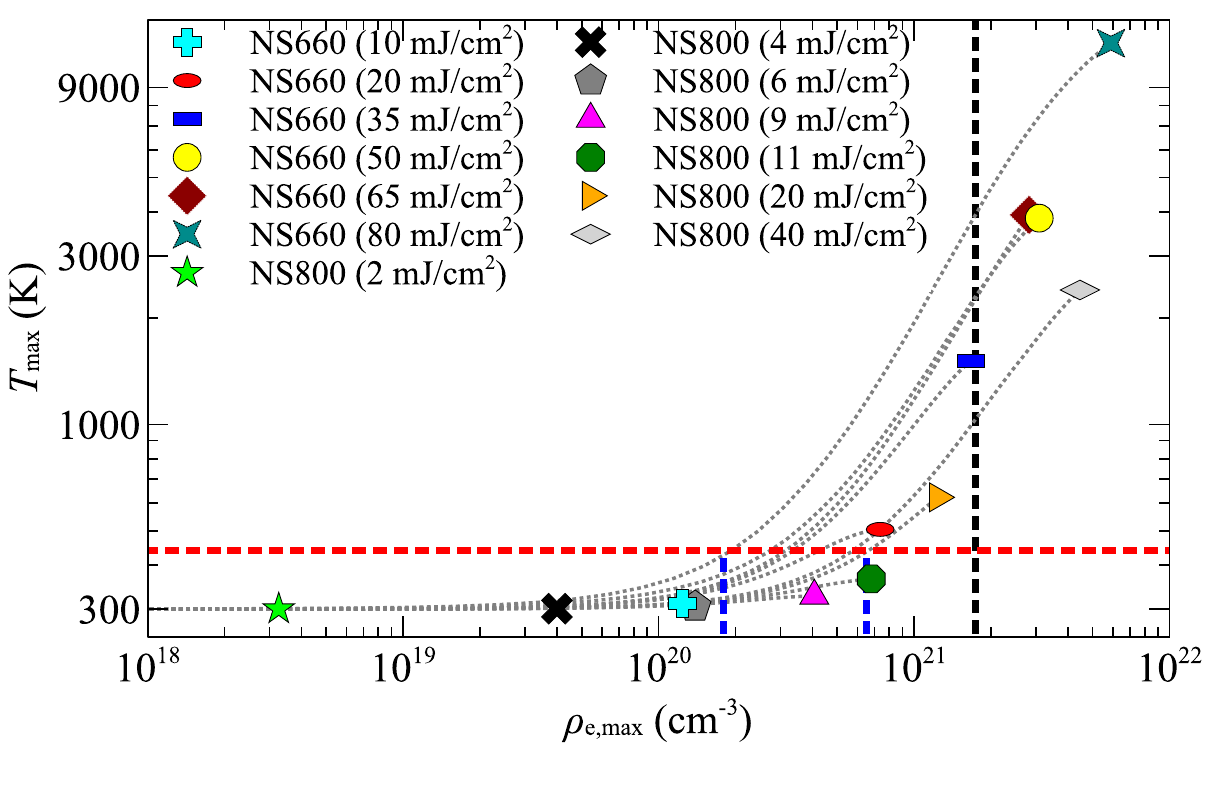}
\caption[Calculated maximum temperature, $T_{\rm max}$, in the vicinity of the gold nanoshell (NS660 and NS800) at the end of the 70~fs pulse]{Calculated maximum temperature, $T_{\rm max}$, in the vicinity of the gold nanoshell (NS660 and NS800) at the end of the 70~fs pulse or when the numerical model stopped versus maximum free electron plasma density, $\rho_{\rm e,max}$, for a given laser fluence. Black dashed line presents the bubble formation threshold due to free electron density generated, $\rho_{\rm bf}~({\rm H_2O})$, ($1.8\times10^{20}~{\rm cm^{-3}}$), while the red dashed line is a threshold of bubble formation due to the temperature, ${\rm T_{bf}}$, (440.7~K) reached in the pure water \cite{Linz2016femto}. The blue line shows an exponential line of best fit, $T_{\rm max}=300{\rm exp}(7\times 10 ^{-22}\rho_{\rm e,max})$ with $R^2=0.943$ for all data points. The grey dashed line represents a new free electron density threshold for bubble formation generated in the vicinity of a nanoparticle, $\rho_{\rm bf}$ (NP).}
\label{fig:TvsRho}
\end{figure}

From the literature review one can find a wide range of free electron plasma densities, $10^{18}-10^{21}$~cm$^{-3}$ \cite{Bloembergen74,Sacchi91,Kennedy95_1,Vogel1996,Feng97,Noack1999,Vogel2005}, that were identified as a parameter that corresponds to the bubble formation and the optical breakdown events. The three orders of magnitude difference in the free electron plasma density coming from the differences in the plasma density calculation and comparison against experimental data. The refinement of this parameter can be done via comparison of the free electron plasma density to more precise parameters such as bubble nucleation temperature and experiment. While the results of the former will be presented below, the latter, a comparison and validation of the model against experimental data on bubble formation was done in our previous publication~\cite{Davletshin2016b}. The maximum temperature, $T_{\rm max}$, of water reached due to free electron thermalization in the vicinity of the gold nanoshell and the maximum free electron density, $\rho_{\rm e,max}$, at the last time step of the optical breakdown modelling in the vicinity of NS660 and NS800 gold nanoshells for different fluences are plotted in Figure~\ref{fig:TvsRho}. It should be noted that not all \gls{fe} solutions reached the end of the 70~fs pulse and some stopped before that. 

In Figure~\ref{fig:TvsRho} the red and black dashed lines represent bubble formation thresholds due to bubble formation temperature, $T_{\rm bf}=440.7~{\rm K}$, and theoretical critical free electron density, $\rho_{\rm crit}=1.74\times10^{21}~{\rm cm^{-3}}$ at 800~nm~\cite{Vogel2005}, respectively. Figure~\ref{fig:TvsRho} shows that for two types of gold nanoshells, NS660 and NS800, at different fluences the $T_{\rm bf}$ is reached at different free electron densities in the range between $\rho_{\rm e,max}=1.75\times10^{20}~{\rm cm^{-3}}$ to  $\rho_{\rm e,max}=6.5\times10^{20}~{\rm cm^{-3}}$ which are marked by blue dashed lines. On the other hand, whenever the $\rho_{\rm crit}$ is reached during different laser fluences, it corresponds to the range of $T_{\rm max}$ between 1000~K to 4000~K, that were reached in the vicinity of the gold nanoshell. Figure~\ref{fig:TvsRho} also clearly shows that with fluence increase, $T_{\rm bf}=440.7~{\rm K}$ is reached with lower $\rho_{\rm e,max}$. Even though, based on the results plotted in Figure~\ref{fig:TvsRho}, one can decrease the three order magnitude uncertainty in the free electron density threshold of bubble formation, the calculation of the boiling temperature, $T_{\rm bf}$, in the vicinity of gold nanoshell should be used for more accurate prediction of \gls{liob}/bubble formation threshold. Contrarily, one will need to estimate a theoretical critical electron plasma density in the vicinity of the nanoparticle in order to have a full picture of \gls{liob} phenomena.

\begin{figure}
\centering
\includegraphics[width=\textwidth]{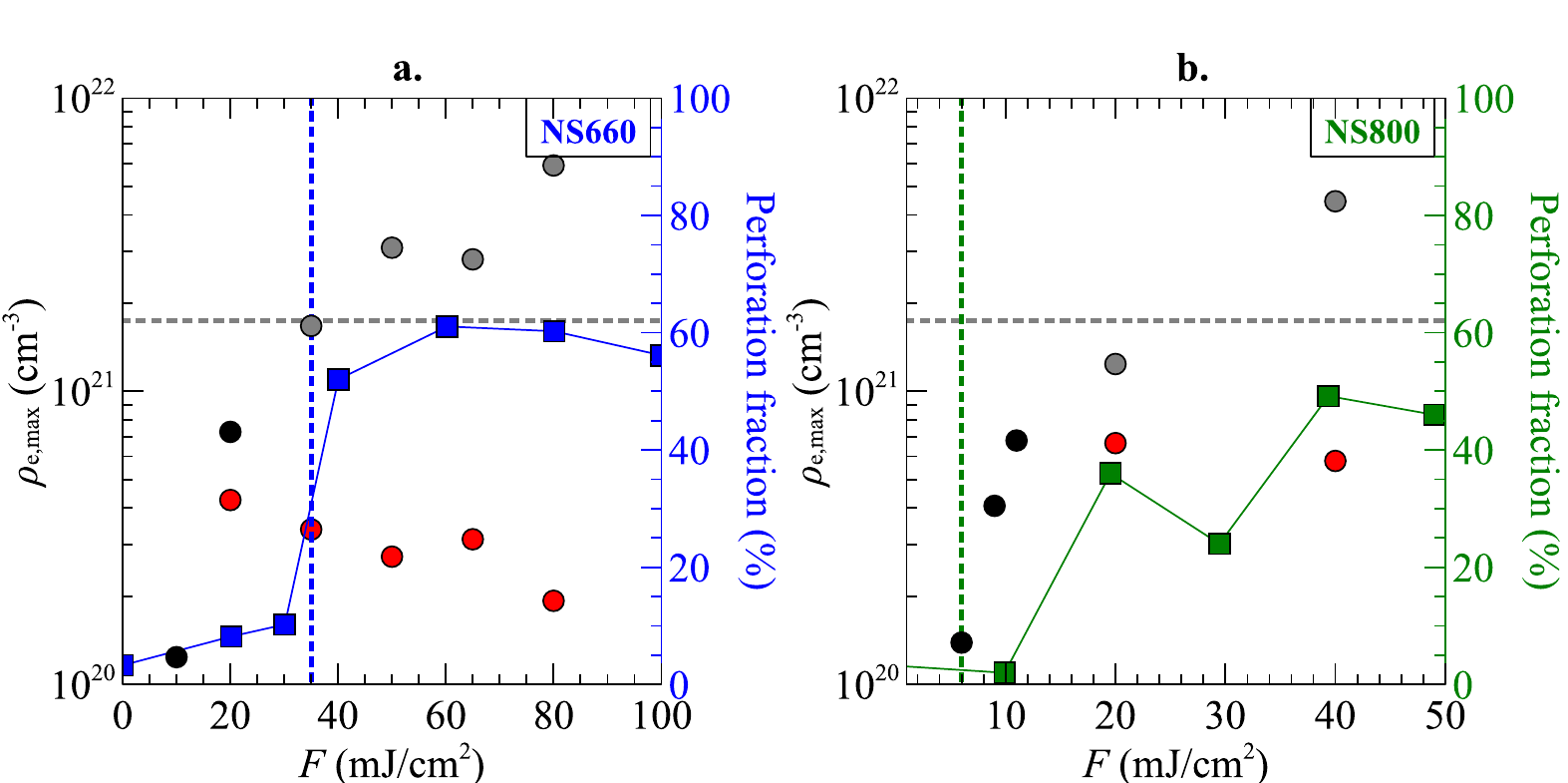}
\caption{A comparison of the maximum free electron density, $\rho_{\rm e,max}$, reached at the end of 70~fs laser pulse (black and grey circles) and calculated in the vicinity of gold nanoshells with the perforation fraction (\%) of the live cells during nanoparticle-mediated \gls{liob} perforation experiment (the blue and green squares), for {\bf a.} NS660 {\bf b.} NS800 gold nanoshells, respectively. The grey circles indicate the simulation fluences that did not run for the entire 70 fs pulse duration. Blue and green dashed lines represent the experimental bubble detection threshold for NS660 and NS800 in pure water, respectively. Grey dashed line represents a theoretical critical free electron density, $\rho_{\rm crit}=1.74\times10^{21}~{\rm cm^{-3}}$ at 800~nm~\cite{Vogel2005}. Red circles represents the maximum free electron density in the vicinity of a nanoshell, that was needed to generate a temperature, $T_{\rm bf}=440.7~{\rm K}$, of a bubble nucleation~\cite{Linz2016femto}.}
\label{fig:Opto_NS}
\end{figure}

The significance of having a full computational understanding of the nanoparticle-mediated \gls{liob} process is explained by comparison of the experimentally measured perforation fraction of live cells during nanoparticle-mediated \gls{liob}~\cite{Lachaine2016} with theoretical predictions. As one can see from \citet{Lachaine2016} in vitro experiments (see Figure~\ref{fig:Opto_NS}), the fraction of the perforated cells is plateauing at laser fluences of $F=40~{\rm mJ/cm^2}$ and $F=20~{\rm mJ/cm^2}$ for NS660 and NS800, respectively. Without doing a theoretical analysis it is hard to understand the plateauing existence and the fluence threshold where it appears. In Figure~\ref{fig:Opto_NS}, the red circles shows $\rho_{\rm e,max}$ at which the $T_{\rm bf}$ was reached for different fluences. It is evident that experimental threshold of bubble formation in pure water (shown as a blue and green dashed lines in Figure~\ref{fig:Opto_NS}) did not correlate with the theoretical prediction of bubble formation due to due to temperature rise to 440.7~K (see red circles in Figure~\ref{fig:Opto_NS}). On the other hand, there is a correlation of reaching boiling temperatures, $T_{\rm bf}$, and fraction of perforated cells. This correlation emphasizes the importance of $T_{\rm bf}$ calculation for prediction of bubble formation threshold  and prediction of cell perforation threshold. More over, $\rho_{\rm e,max}$ calculations will have uncertainties and can not be related to a single density value for the perforation threshold during nanoparticle-mediated optical breakdown experiments. Furthermore, the existence of the plateau is explained by generation of $\rho_{\rm e,max}$ over theoretical critical free electron density, $\rho_{\rm crit}$, during the laser pulse duration so that medium becomes highly reflective and the absorption cross-section of the plasma decreases significantly (see grey dashed line and grey circles in Figure~\ref{fig:Opto_NS}). Hence, the computation of both parameters, $T_{\rm bf}$ and $\rho_{\rm crit}$ can accurately predicts the outcome of the nanoparticle-mediated \gls{liob} live cell perforation experiments.

\section{Conclusion}

In conclusion, this study demonstrates the importance of theoretical prediction of temperature and the free electron density rise in the vicinity of gold nanoshell for the outcomes of the live cell perforation experiments. While theoretical prediction of reaching bubble formation temperatures has a good agreement with the threshold of live cells perforation, the calculation of theoretical critical free electron density can identify perforation regime where the increase of the laser fluence will have little effect on perforation fraction. 

% \begin{acknowledgement}

% Please use ``The authors thank \ldots'' rather than ``The
% authors would like to thank \ldots''.

% The author thanks Mats Dahlgren for version one of \textsf{achemso},
% and Donald Arseneau for the code taken from \textsf{cite} to move
% citations after punctuation. Many users have provided feedback on the
% class, which is reflected in all of the different demonstrations
% shown in this document.

% \end{acknowledgement}

% \begin{suppinfo}

% This will usually read something like: ``Experimental procedures and
% characterization data for all new compounds. The class will
% automatically add a sentence pointing to the information on-line:

% \end{suppinfo}

\bibliography{OSA_style}

\end{document}